\documentstyle[12pt]{article}
\begin{document}
\newsavebox{\uuunit}
\sbox{\uuunit}
    {\setlength{\unitlength}{0.825em}
     \begin{picture}(0.6,0.7)
        \thinlines
        \put(0,0){\line(1,0){0.5}}
        \put(0.15,0){\line(0,1){0.7}}
        \put(0.35,0){\line(0,1){0.8}}
       \multiput(0.3,0.8)(-0.04,-0.02){12}{\rule{0.5pt}{0.5pt}}
     \end {picture}}
\newcommand {\unity}{\mathord{\!\usebox{\uuunit}}}
\newcommand{\half}{{\textstyle\frac{1}{2}}}
\newcommand{\for}{{\textstyle\frac{1}{4}}}
\newcommand{\eqn}[1]{(\ref{#1})}
\newcommand{\dr}{\raise.3ex\hbox{$\stackrel{\leftarrow}{\partial }$}}
\newcommand{\dl}{\raise.3ex\hbox{$\stackrel{\rightarrow}{\partial}$}}
\newcommand{\cR}{{\cal R}}
\begin{titlepage}
\begin{flushright} CERN-TH.6541/92\\ KUL-TF-92/24\\ hepth@xxx/9206097
\end{flushright}
\vfill
\begin{center}
{\large\bf The regularized BRST Jacobian of pure Yang-Mills theory}
   \\
\vskip 5.mm
{\bf F. De Jonghe$^1$, R. Siebelink$^2$, W. Troost$^3$,
S. Vandoren$^4$}\\
Instituut voor theoretische fysica
\\K. U. Leuven,
B-3001 Leuven, Belgium
\\[0.3cm]

{\bf P. van Nieuwenhuizen$^5$} and {\bf A. Van Proeyen$^6$} \\
Theory Division, CERN\\ CH-1211 Geneva 23, Switzerland
\end{center}
\begin{quote}
The Jacobian for infinitesimal BRST transformations of path integrals
for pure Yang-Mills
theory, viewed as a matrix $\unity +\Delta J$ in the space of
Yang-Mills fields and (anti)ghosts, contains off-diagonal terms.
Naively, the trace of $\Delta J$ vanishes, being proportional to the
trace of the structure constants. However, the consistent regulator $\cR$,
constructed from a general method, also contains off-diagonal terms.
An  explicit computation  demonstrates that the regularized
Jacobian $Tr\ \Delta J\exp -\cR /M^2$ for $M^2\rightarrow \infty $ is
the variation of a local counterterm, which we give. This is a direct
proof at the level of path integrals that there is no BRST anomaly.
\hrule width 5.cm
{\small\small
\noindent $^1$ Aspirant, NFWO, Belgium; Bitnet FGBDA16 at BLEKUL11\\
\noindent $^2$
Aspirant, NFWO, Belgium; Bitnet FGBDA04 at BLEKUL11\\
\noindent $^3$
Bevoegdverklaard navorser, NFWO, Belgium; Bitnet FGBDA19 at BLEKUL11\\
\noindent $^4$
Bitnet FGBDA43 at BLEKUL11\\
\noindent $^5$
On leave from the Institute for Theoretical Physics,
SUNY at Stony Brook, NY 11794, USA; bitnet VANNIEU
at SUNYSBNP\\
\noindent $^6$ On leave from Instituut voor theoretische fysica, K.U.
Leuven;
Onderzoeksleider, N.F.W.O. Belgium;
Bitnet FGBDA19 at BLEKUL11}
\normalsize
\end{quote}
\begin{flushleft} CERN-TH.6541/92 \\ KUL-TF-92/24 \\
June 1992
\end{flushleft}
\end{titlepage}
\setlength{\textheight}{522pt}
The rigid \footnote{For local (gauged) BRST symmetry, see F.J. Ore
and P. van Nieuwenhuizen, Nucl. Phys. {\bf B204} (1982) 317.}
 BRST symmetry
\cite{BRST} of the quantum action for gauge theories has become
an essential tool in the path-integral approach. One uses the BRST
symmetry to obtain Ward identities \cite{Ward}, which are then used
to prove perturbative renormalizability \cite{renorm} and unitarity
\cite{unit}.
In principle, the Ward identities can contain an anomaly which appears
in the path-integral approach as a deviation from unity of the
Jacobian for BRST transformations. It is often argued that there is
no anomaly in the BRST Ward identities because $\Delta J$ is
proportional to the trace of the structure constants, which vanishes
for semi-simple gauge groups.
However, this argument, though widely repeated and presented in various
textbooks, is patently false. For example, the same line of reasoning
would conclude that there never are chiral anomalies since $tr\,\gamma
_5=0$. For chiral anomalies one knows that after regularization with
a regulator $\cR $, the regularized trace
$tr\,\gamma _5 \exp -\cR /M^2$ for $M\rightarrow \infty $ no
longer vanishes in general and yields the chiral anomaly.
For chiral symmetries, the choice of $\cR$ is not very important, as
``[one] can show that under broad assumptions the one loop anomalies
depend only on the quantum numbers of the elementary fields, and not
on the specific Lagrangians chosen" \cite{AGW}. This is due to the
topological nature of the chiral anomaly. In general,
the choice of $\cR$ clearly matters. For
example, the Jacobian for Weyl symmetry is field-independent, and if
one were to choose a regulator which is also field-independent, one could
never obtain anomalies proportional to the curvatures.
Recently, a general method was developed for constructing a consistent
regulator for the measure of any quantum field theory \cite{anomPV}.
This regulator is equivalent to Pauli-Villars regularization of the
Feynman diagrams of the effective action, and yields anomalies
which satisfy the consistency conditions \cite{WZcc}. It is this
regulator which we will use to compute the BRST anomaly.
For BRST symmetry it is computationally not
obvious that the regularized trace $Tr
\Delta J\exp -\cR /M^2$ still vanishes in the limit
$M\rightarrow \infty$, although one would
expect so, given that indirect (cohomological) arguments say so
\cite{coho}.
In this article we will demonstrate by a direct calculation that the
BRST anomaly in the Ward identities indeed vanishes
\footnote{For convenience
we will use the word `anomaly' for this regulated trace, even when it
is the variation of a local counterterm and can be removed.}.
\newpage

The derivation of the Ward identities is based on the
Shakespeare theorem~\footnote{
\begin{verse}
... \^{o} be some other name.\\
Whats in a name? that which we call a rose,\\
By any other word would smell as sweete,\\
So Romeo would were he not Romeo cald, \\
Retaine that deare perfection...
\cite{Shakesp}
\end{verse}}
 according to which one may rename
the integration variables $\phi ^j$ everywhere (in the measure, in the
action and in the coupling to external sources) by $\phi '^j=\phi ^j +
\delta \phi ^j$
\begin{eqnarray}
Z(J)&=&\int {\cal D}\phi ^j\,\exp {\textstyle{\frac{i}{\hbar }}}\left[
S_{qu}(\phi )+ \int J_j\phi ^j d^4x\right] \nonumber\\
&=&\int {\cal D}\phi'^j\,\exp {\textstyle{\frac{i}{\hbar }}}\left[
S_{qu}(\phi')+ \int J_j\phi'^j d^4x\right]\ .
\end{eqnarray}
We take for $\delta \phi ^j$ an infinitesimal BRST transformation.
The quantum action $S_{qu}(\phi )$ consists of an $\hbar =0$ part
which is BRST invariant, and
possibly local counterterms $\hbar M_1$, where $M_1$ is in general
a power series in $\hbar $.
This leads to the formal Ward identity
\begin{equation}
\bigg< Tr\,\Delta J +  i\delta M_1 +
{\textstyle{\frac{i}{\hbar }}}
 \int J_j\delta \phi ^j d^4x \bigg>=0     \label{eq:Ward}
\end{equation}
where
\begin{equation}
\Delta J^j{}_k=\frac{\dr\delta\phi^j}{\partial \phi^k}
\end{equation}    (with right derivatives)
is the deviation of the Jacobian matrix from unity.
We shall discuss the
regularization of the trace $Tr\,\Delta J$ in pure Yang-Mills theory,
with $\phi ^j$ equal to the fields $b$ (antighosts), $Q_\mu $ (quantum
Yang-Mills fields) and $c$ (ghosts). However, for reasons to be
explained, we shall use an action containing an extra, external, gauge
field $B_\mu $, which interpolates between ordinary quantum field theory
and the background field method.
This action $S=S_{qu}(\hbar =0)$
reads
\begin{equation}
S=Tr\int d^4x\left\{ -\textstyle{\frac{1}{4}}F_{\mu \nu }(Q)^2
-\textstyle{\frac{1}{2}}\left[
 D^\mu (B)(Q_\mu -B_\mu )\right] ^2
- D_\mu (B) b\cdot D^\mu (Q) c
 \right\}   \label{SYMB}
\end{equation}
where
$D_\mu (X)Y=\partial _\mu Y +[X_\mu ,Y]$ and the trace $Tr$ is over
gauge indices.
For $B_\mu=0$, we obtain the Feynman gauge, while for $B_\mu \neq 0$
we recognize the action for the background field formalism (after
shifting $Q_\mu $ to $Q_\mu +B_\mu $). This
last action is invariant under two symmetries
\begin{enumerate}
\item local background symmetry, under which both the background field
$B_\mu$ and the quantum field $Q_\mu $ transform as gauge fields, and
$b$ and $c$ as vectors
\begin{eqnarray}
\delta Q_{\mu} &=& D_\mu (Q) \lambda  \nonumber\\
\delta B_\mu &=& D_\mu (B) \lambda  \nonumber\\
\delta b&=& [b,\lambda ]   \nonumber\\
\delta c&=& [c,\lambda ]\ .\label{bgrtr}
\end{eqnarray}
with $\lambda $ the local commuting Lie-algebra valued Yang-Mills
parameter.
\item rigid BRST symmetry, under which the background field is inert
\begin{eqnarray}
\delta Q_\mu &=& D_\mu (Q) c\Lambda \nonumber\\
\delta B_\mu &=& 0  \nonumber\\
\delta b &=&  - D^\mu (B) (Q^{\mu} - B^{\mu})\Lambda \nonumber\\
\delta c &=&\half [c,c] \Lambda  \label{BRSTtr}
\end{eqnarray}
with $\Lambda $ the constant, anticommuting BRST parameter.
\end{enumerate}
In this article we shall discuss the Jacobians for these symmetries,
as they appear in the path-integrals in \eqn{eq:Ward}.

If one (erroneously) neglects regularization, one would conclude that
the trace of $\Delta J$ in \eqn{eq:Ward} vanishes for both symmetries
in \eqn{bgrtr} and \eqn{BRSTtr}, as the diagonal entries are
proportional to $f^a{}_{ab}$ in each case (where
$f^a{}_{bc}$ are the structure constants of a Lie algebra, which are
traceless for the semi-simple groups which we consider).
As we explained above this conclusion is incorrect.

In view of the importance of BRST symmetry, we think that a direct,
explicit calculation of the anomaly (the trace of $\Delta J
\exp -\cR /M^2$)
for both symmetries is useful. Of course we shall discuss how
to determine $\cR $.
In addition, we shall determine by direct computation
whether the anomaly can be written
as the variation of a local counterterm (and thus removed)
and when it actually vanishes.

We work in the generalized gauge with $B_\mu $ present in order to
avoid an accidental vanishing of the anomaly. To draw a comparison with
string theory, we prefer to work in a `general gauge' (like
$g_{\alpha\beta }=G_{\alpha\beta }$ where $G_{\alpha \beta }$ is an
arbitrary background field) rather than a special gauge. If one were to
choose the
gauge $g_{\alpha\beta }=\eta _{\alpha \beta }$, the anomaly
${\cal A}= c\sqrt{g}R(g)=0$ would seem to vanish.
Of course, this is not an allowed gauge since it cannot be reached
using the gauge symmetries without anomalies. Similarly,
in our case, we still have
to prove that there are no Yang-Mills anomalies, so we should not use the
Yang-Mills symmetry to
choose a special gauge. In the Batalin-Vilkovisky field-antifield formalism
\cite{BV} there is a natural way to work with general gauges
 \cite{bvsb}. Usually one
eliminates antifields by putting
$\phi ^*_j=\frac{\partial\Psi }{\partial\phi^j}$, where $\Psi $ is a
`gauge fermion'.
This is equivalent to
first making a canonical transformation
with a `generating function' $\Psi $, and then projecting onto the
hypersurface $\phi ^*_j=0$. However, if one does not project
onto this hypersurface and keeps the antifield dependence,
then these play the role of arbitrary gauge parameters.
For the string e.g., the choice $\Psi =b^{\alpha \beta }(g_{\alpha
\beta} -\eta _{\alpha\beta})$ yields
$g_{\alpha \beta }=\eta
_{\alpha \beta }-b^*_{\alpha \beta }$, where the last term is the
antifield of the antighost.
Clearly, whether the anomaly is expressed as a function of a general
background metric $G_{\alpha \beta }$, or as a function of the
$b^*_{\alpha \beta }$-antifield is the same thing.
We have done the calculations also with antifields (for $B_\mu=0$, which
is then sufficient), but for
simplicity, we shall present here the calculations with
\eqn{SYMB} (and zero antifields), as this action is of practical interest
and general enough for our purposes.
\vspace{0.5cm}

The most practical method for regularisation of Feynman diagrams is
the dimensional regularisation method. It keeps gauge invariance
at all stages when there are no $\gamma _5$ or similar
dimension-dependent objects present. Therefore there are no anomalies in
this regularization scheme in these cases.
However, dimensional regularization can not be applied directly
to the path integral measure.
As shown by Fujikawa \cite{Fujikawa},
in path integrals the anomalies come from the measure. However, in the
original works, it was not known which regulators would give
consistent \cite{WZcc} anomalies. This problem was solved in
\cite{anomPV}, where a general recipe was obtained which yields a
consistent regulator for any quantum field theory, once the quantum
action is given.
Input for this method is a mass matrix $\phi ^iT_{ij}
\phi ^j$, which must be
 non-singular.
Output is a consistent regulator
$\cR $, given by
\begin{equation}
\cR ^i\,_j = (T^{-1})^{ik}S_{kj} \ ;\qquad
S_{ij}=\frac{\dl }{\partial\phi^i}\frac{\dr}{\partial \phi^j}S\ .
\end{equation}
The regularized trace (really a supertrace due to the presence of the
\hfil\break (anti)ghosts, hence denoted by $str$) is then given by
\begin{eqnarray}
{\cal A}&=&\lim_{M^2\rightarrow \infty }
str\ \Delta J \exp (T^{-1}S/M^2)\nonumber\\
&=&\lim_{M^2\rightarrow \infty }
str \Delta J_s e^{T^{-1}S/M^2}\ ,\label{anomJs}
\end{eqnarray}
with
\begin{equation}
\Delta J_s\equiv \half \left(
\Delta J + T^{-1} \Delta J^t T \right)\ .
\end{equation}
The transposition on $\Delta J$ in this equation refers to a
supertransposition
\footnote{All our matrices are supermatrices of bosonic type. Then
the supertranspose for matrices with 2 lower indices, as $S$ and $T$,
 is defined by
$(T^t)_{ij}= T_{ji} (-)^{i+j+ij}$ where $(-)^i$ is $+$ for entries
related to $Q_\mu$ and $-$ for entries related to $b$ and $c$. This rule
follows from the definition that
$\phi ^i (T^t)_{ij}\varphi ^j = \varphi ^j T_{ji}\phi ^j$. For matrices
with two upper indices, we have $(M^t)^{ij}=M^{ji}(-)^{ij}$, as can be
derived for $T^{-1}$ if we impose
$\left( T^{-1}\right) ^t  T^t=\unity $. Finally for matrices with
mixed indices like $J^i{}_j$ the supertransposition rule
follows from the product of matrices of the previous types, imposing
$(MT)^t=T^tM^t$. We have $(J^t)_i{}^j=(-)^{i(j+1)} J^j{}_i$. For these
matrices the supertrace is $str\ J= (-)^i J^i{}_i$ which has the
property that $str\ J^t=str\ J$ and
$str\ AB= str\ BA$. These properties, with the knowledge that $S$ and
$T$ are super-symmetric by their definition, lead to the equality in
\eqn{anomJs}.}
 of the $\Delta J$ matrix, a transposition of the
derivative operators which amounts to an extra minus sign after partial
integration, and a transposition of the Lie-algebra representation matrices
which amounts to another minus sign (in the adjoint representation).
This rewriting of
${\cal A}$ makes use of the super-symmetry of $S_{ij}$ and $T_{ij}$, and
will simplify calculations later on.

It is conjectured
(and checked in many examples) that\\
- this method gives a consistent anomaly, i.e.,
 the Wess-Zumino conditions
are satisfied.\\
- the expression for $\cal A$ is gauge dependent, but this
  dependence can be absorbed in a counterterm.\\
- different choices of the mass term lead to different expressions of
${\cal A}$ which again differ by the variation of a local counterterm.\\
This example will provide another check on the first two of
these conjectures.

We choose a mass term which is invariant under rigid
Yang-Mills transformations~:
$tr\ \int d^4x\left[ Q_\mu Q^\mu +2bc\right]$.
We write the matrix-entries in order of decreasing ghost number,
namely in the order $b$, $Q^\mu $, $c$ : this makes
the triangular nature of the
matrices to follow more manifest. Then
\begin{equation}
T=\left(  \begin{array}{ccc}
  0 & 0 & 1 \\
  0 & \eta _{\mu\nu }& 0 \\
  -1 & 0 & 0     \end{array}  \right)\ ;\qquad
T^{-1}=\left(  \begin{array}{ccc}
  0 & 0 &- 1 \\
  0 & \eta ^{\mu\nu }& 0 \\
  1 & 0 & 0     \end{array}  \right)\ .
\end{equation}

For the local background symmetry, $\Delta J$ is diagonal, with entries
$f^a{}_{bc}\lambda ^c$,
$f^a{}_{bc}\lambda ^c\delta ^\mu _\nu $ and
$f^a{}_{bc}\lambda ^c$, respectively. One may verify that $\Delta J_s$
vanishes in this case. Hence, the background symmetry is preserved
at the quantum level in an almost obvious way.

For the rigid BRST symmetry, to which we devote the rest of this
article, we find that $\Delta J$ is off-diagonal and contains derivatives,
\begin{eqnarray}
\Delta J=\left(\begin{array}{ccc}
0 & -D_{\nu}(B) & 0\\
0 & -c\delta ^\mu\, _\nu  &-D^\mu (Q) \\
0 & 0 &-c
\end{array}\right)(x)\delta (x-y)\Lambda \ ,
\end{eqnarray}
where entries such as $c$ stand for the matrices $f^a{}_{cb}c^c$.
The symmetrized $\Delta J_s$ contains no derivatives but is
purely algebraic:
\begin{equation}
\Delta J_s = \frac{1}{2}\left(\begin{array}{ccc}
c & Q_\nu -B_\nu & 0\\
0 & 0 & -Q^\mu +B^\mu \\
0 & 0 & -c
\end{array}\right)(x)\delta (x-y)\Lambda \ .
\end{equation}
This fact will greatly simplify the evaluation of the supertrace in
\eqn{anomJs}.
Furthermore, the operator matrix $T^{-1}S$ is given by
\begin{equation}
T^{-1}S=\left(\begin{array}{ccc}
               D_\alpha (Q) D^\alpha (B) &-(D_\nu (B) b) & 0\\
 c D^\mu (B) &  R^\mu {}_\nu & (D^\mu (B) b)\\
 0 & -D_\nu (B) c & D_\alpha (B) D^\alpha (Q)
                 \end{array}\right)(x)\delta (x-y)\ ,
\end{equation}
where
\begin{equation}
R^\mu {}_\nu =
 D_\alpha(Q)D^\alpha(Q) \delta ^\mu_\nu - D^\mu (Q)D_\nu (Q)
   + D^\mu (B)D_\nu (B) + 2 F^\mu{}_\nu(Q)
\end{equation}
and the covariant derivatives act as far as $\delta ^4(x-y)$,
unless put within explicit brackets. This expression can be cast
into the form
\begin{equation}
T^{-1}S=
\left( \partial_{\alpha}\unity +{\cal Y_{\alpha}} \right)\eta ^{
\alpha \beta }
\left( \partial_{\beta}\unity +{\cal Y_{\beta}} \right)  + E
\label{regulcast}
\end{equation}
where $\unity $ and $\cal Y_\alpha $ are $6\times 6$ matrices
with entries in the
adjoint representation of the Yang-Mills Lie algebra, and
$\alpha,\,\beta $ are ordinary Minkowski
indices. The derivatives in (\ref{regulcast}) are explicit, i.e., ${\cal
Y_{\alpha}}$ and $E$ do not contain free
derivative operators any more. We find
the following expression for $\cal Y_{\alpha}$ and $E$ in $d$ dimensions
\begin{eqnarray}
 {\cal Y_{\alpha}}&=& Q_\alpha \unity + \frac{1}{2}
\left( \begin{array}{ccc}
      -Q'_{\alpha}  & 0 & 0 \\
       c \delta ^\mu{}
        _\alpha &-Q'^\mu\eta_{\nu\alpha}-Q'_\nu \delta ^\mu{}
_\alpha       & 0 \\
        0 & -c \eta_{\alpha\nu} & -Q'_\alpha
     \end{array} \right)\nonumber\\
 E&=& -\frac{1}{4}Q'^2 \unity +
\left( \begin{array}{ccc}
      - \half D_\alpha (B)Q'^\alpha  & -D_\nu(B) b & 0 \\
V^\mu &E^\mu{} _\nu          & D^\mu(B) b \\
{\textstyle\frac{d}{4}}c^2 &   -V_\nu ^T &\half D_\alpha (B)Q'^\alpha
     \end{array} \right)
\end{eqnarray}
where $Q'^{\nu}= Q^{\nu} - B^{\nu}$ and\footnote{Antisymmetrization
$[\mu\nu]$ is done with weight 1, i.e., $\half (\mu \nu -\nu \mu )$.}
\begin{eqnarray}
V^\mu &=&-\half D^\mu(B) c+\for((d-1)Q'^\mu c-cQ'^\mu )\nonumber\\
-V_\nu ^T &=&
     -\half D_\nu(B) c+\for( Q'_\nu c-(d-1)cQ'_\nu)\nonumber\\
E_{\mu \nu }&=&
2F_{\mu\nu}(B) +
3D_{[\mu }(B)Q'_{\nu ]} +({\textstyle \frac{3}{2}-\frac{d}{4}})
Q'_\mu Q'_\nu-Q'_\nu Q'_\mu
\end{eqnarray}

The trace in \eqn{anomJs} can be evaluated using the heat kernel. As long
as $\Delta J_s$ is algebraic, i.e., contains no derivative operator
(which is the case for all applications we consider
in this paper), only the value of this kernel at coincident points is
needed. In the limit of large
$M^2$ it can be calculated by a variety of methods. We read off the result
from \cite{Gilkey}, generalized to the mixed bosonic and fermionic case.
We are interested in
the terms independent of $M^2$. In four dimensions
they are usually denoted by $a_2$ and
read
\begin{equation}  a_2=
\frac{1}{(4\pi)^2}  \left(
 \frac{1}{12}  W_{\alpha \beta} W^{\alpha \beta} +
\frac{1}{2}  E^{2} +
 \frac{1}{6}\Box E \right) \ ,
\end{equation}
where
\begin{eqnarray}
W_{\alpha \beta}&=&\partial _\alpha {\cal Y}
_\beta -\partial _\beta {\cal Y}_\alpha
+[{\cal Y}_\alpha,{\cal Y}_\beta]\nonumber\\
\Box E&=&\nabla _\alpha \nabla ^\alpha E \nonumber\\
\nabla_\alpha X&=&\partial _\alpha X + [{\cal Y}_\alpha ,X]\ .
\end{eqnarray}
In two dimensions they are denoted by $a_1$ and read
\begin{equation}
a_1= \frac{1}{4\pi } E\ .
\end{equation}

As a check on these results we compute the trace anomalies. They are
obtained by taking the trace of the Weyl Jacobian $J_W$ with the
$a_n$ coefficients. For two dimensions the Weyl weights of $c$, $b$ and
$Q_\mu$ are all equal,
see \cite{BastPvN}, hence the
Weyl anomaly becomes proportional to the trace of $E$ ($d=2$). This
vanishes, in agreement with the fact that the trace anomaly for spin 1
fields in $d=2$ is zero. For $d=4$ one has $J_W=\mbox{diag}
(0,\half \sigma (x),\sigma (x))$, see
\cite{BastPvN}, yielding
the correct result. Another check on the correctness of $a_2$
(except the $\Box E$ term) is that after integration over space-time,
they yield the one-loop counterterms, deduced by dimensional
regularization and Feynman diagrams in \cite{GtHbf}.

As we already mentioned,
in the Yang-Mills case $\Delta J_s$ is also algebraic, due to the
symmetrization in \eqn{anomJs}. The
anomaly is obtained by computing the supertrace
\begin{equation}
{\cal A} =
\frac{1}{(4\pi)^2}
 str\ \Delta J_s
\left(
 \frac{1}{12}  W_{\alpha \beta} W^{\alpha \beta} +
\frac{1}{2}  E^{2} +
 \frac{1}{6}\Box E \right) \ . \label{AnGilk}
\end{equation}
The problem of obtaining the BRST anomaly for pure Yang-Mills theory
is thus reduced to the evaluation of the supertrace in \eqn{AnGilk}.
The result reads
\begin{eqnarray} {\cal A}=
\frac{1}{(4\pi )^2}\frac{1}{12}
tr\ [D^\nu(B) c]
\left[ 4 Q'_\mu Q'_\nu Q'^\mu -8 Q'^\mu D_{[\mu}(B)Q'_{\nu
]} -4Q'_\nu D_\mu(B)Q'^\mu  \right.\nonumber\\
\left. +D_\mu(B) D^\mu(B) Q'_\nu  -3
D_\nu(B) D_\mu(B) Q'^\mu \right]\ . \end{eqnarray}
If this is to be a consistent anomaly, its BRST variation should vanish.
Indeed, it does. Expecting that there is no genuine BRST anomaly, this
expression is expected to be the BRST variation of a local
counterterm. Indeed, it is: ${\cal A}=\delta M_1$, with
\begin{eqnarray}
M_1  =
\frac{1}{(4\pi )^2}\frac{1}{12} tr\  \left[
{\textstyle
\frac{3}{2}}(D_\mu(B)Q'^\mu)^2 - \half
(D_\mu(B)Q'_\nu )(D^\mu(B) Q'^\nu )    \right.
\nonumber\\ \left.
 -2 Q'^\mu(D_\mu (B)Q'_\nu )Q'^\nu  +{\textstyle
 \frac{3}{2}}Q'_\mu Q'_\nu Q'^\mu Q'^\nu  - \half Q'^2 Q'^2\right]\ .
\end{eqnarray}
For computations, a suitable alternative form is given by
\begin{eqnarray}
M_1=
\frac{1}{(4\pi )^2}\frac{1}{12} tr\
\left[  \textstyle{\frac{3}{2}}(D_\mu(B) Q'^\mu
)^2-\textstyle{\frac{1}{2}}(D_\mu(B)Q'_\nu )(D^\nu(B) Q'^\mu )
\right.\nonumber\\ \left.
 -(D_\mu(B)Q'_\nu )(D^\mu(B) Q'^\nu )
+Q'_\mu Q'_\nu Q'^\nu Q'^\mu +\textstyle{\frac{1}{4}}F'_{\mu \nu
}F'^{\mu \nu }\right) \label{M1c}
\end{eqnarray}
where
\begin{equation}
F'_{\mu \nu }= D_\mu(B)
 Q'_\nu -D_\nu(B) Q'_\mu +[Q'_\mu , Q'_\nu ]=F_{\mu \nu
}(Q)-F_{\mu \nu }(B). \end{equation}
In the background field formalism $Q_\mu $ gets replaced by an external
field $A_\mu $, and then one usually chooses the two external fields
equal ($B_\mu =A_\mu $). In this case the anomaly vanishes without
having to invoke a counterterm. In ordinary field theory one puts
$B_\mu =0$; in this case one needs a nontrivial counterterm $M_1$.

In conclusion, we have seen that within the framework of
path-integrals, one can give a direct and complete derivation of the
BRST Ward identities, on which perturbative unitarity and
renormalizability are based. The precise form of the local
counterterm $M_1$ which must be added to the quantum action in order
that the BRST anomaly cancels, depends of course on the
regularization scheme used to compute the effective action.
If one uses Pauli-Villars regularization with the given mass term,
$M_1$ is given in \eqn{M1c}. If one uses dimensional regularization,
one has $M_1=0$. However, the precise form of $M_1$ is not needed in
general.

\end{document}